\documentclass[twocolumn,pra, showpacs, showkeys]{revtex4}
\usepackage{graphicx}
\usepackage{dcolumn}
\usepackage{bm}

\begin{document}

\preprint{APS/123-QED}

\title{An effective quasi-one-dimensional description of a spin-1 atomic
condensate}
\author{Wenxian Zhang and L. You}
\address{School of Physics, Georgia Institute of Technology, Atlanta GA
30332-0430, USA}
\date{\today}

\begin{abstract}
Within the mean field theory we extend the effective quasi-one-dimensional
nonpolynomial Schr\"{o}dinger equation (NPSE) approach to the description of a
spin-1 atomic condensate in a tight radial confinement geometry for both weak
and strong atom-atom interactions. Detailed comparisons with full time
dependent 3D numerical simulations show excellent agreement as in the case of a
single component scalar condensate, demonstrating our result as an efficient
and effective tool for the understanding of spin-1 condensate dynamics observed
in several recent experiments.
\end{abstract}



\maketitle
\narrowtext


Although our ability to perform numerical simulations keeps increasing with
computer technology, full 3D time dependent calculations still represent a
significant challenge. In many situations, one explores the inherent system
symmetries, e.g., cylindrical and spherical symmetries in space, to reduce the
number of spatial dimension from 3D to 2D or even 1D. The description of atomic
condensate dynamics in terms of a mean field theory is such an example. With a
tight radial confinement, a condensate becomes cigar-shaped. Several effective
1D approaches have been developed \cite{Chiofalo00, Gerbier04, Salasnich04},
with the simplest of them assuming a fixed transverse Gaussian profile. Recent
studies, however, have indicated that the effective quasi-one-dimensional (1D)
nonpolynomial Schr\"odinger equation (NPSE) is the most powerful and efficient
tool, at least for a weakly interacting atomic condensate \cite{Salasnich04}.
In this brief report, we generalize such an NPSE approach to the case of a
spin-1 atomic condensate in a cigar-shaped trap.

Multi-component atomic condensates or spinor condensates have become an
actively investigated topic in atomic quantum gases \cite{Myatt97,
Stamper-Kurn98, Stenger98, Barrett01, Chang04, Chang04a, Schmaljohann04,
Kuwamoto04, Ho98, Law98, Pu99, Yi02, Pu00, NJP}. The examples that have been
experimentally realized include that of two-component pseudo-spin-1/2
\cite{Myatt97}, three-component spin-1,  and five-component spin-2 condensates
\cite{Stamper-Kurn98, Stenger98,Chang04a,Kuwamoto04}. Several recent
experiments have observed interesting coherent spatial fragmentation of the
spin-1 condensate when it is confined in a single running wave optical trap,
i.e., in a cigar-shaped trap. To provide a proper theoretical description for
these observations, numerical approaches have been used to study the nonlinear
spatial-temporal dynamics for a spin-1 condensate. It is therefore desirable to
have a more efficient theoretical approach instead of the 3D coupled
Gross-Pitaevskii (GP) equations that is uniformly valid to both strongly and
weakly interacting limits.

A spin-1 Bose condensate is described by the Hamiltonian in second quantized
form (repeated indices are summed) as \cite{Ho98}
\begin{eqnarray}
{\cal H} &=&\int d\vec r\, \left[ \Psi_i^\dag\left(-{\hbar^2 \over 2M}\nabla^2
+ V_{\rm ext}+E_i\right) \Psi_i \right. \label{h} \\
&+&\left. {\frac{c_0}{2}} \Psi_i^\dag\Psi_j^\dag\Psi_j\Psi_i +{c_2\over 2}
\Psi_k^\dag\Psi_i^\dag\left(F_\gamma
\right)_{ij}\left(F_\gamma\right)_{kl}\Psi_j \Psi_l\right], \nonumber
\end{eqnarray}
where $\Psi_j(\vec r)$ ($\Psi_j^\dag$) is the field operator that annihilates
(creates) an atom in the $j$th internal state at location $\vec r$, $j=+,0,-$
denotes atomic hyperfine state $|F=1, m_F=+1,0,-1\rangle$, respectively. $M$ is
the mass of each atom and $V_{{\rm ext}}(\vec r)$ is an
internal-state-independent trap potential. Terms with coefficients $c_0$ and
$c_2$ of Eq. (\ref{h}) describe elastic collisions of two spin-1 atoms,
expressed in terms of the scattering lengths $a_0$ ($a_2$) in the combined
symmetric channel of total spin $0$ ($2$), $c_0=4\pi\hbar^2(a_0+2a_2)/3M$ and
$c_2=4\pi\hbar^2(a_2-a_0)/3M$. $F_{\gamma=x,y,z}$ are spin-1 matrices
\cite{NJP}. Since an external magnetic field is usually present in experiments,
it is also included in our formulation. For simplicity, the magnetic field $B$
is taken along the quantization axis ($\hat z$). The Zeeman shift on each
atomic state is then given by the Breit-Rabi formula \cite{para1}.

Adopting the mean field theory when the condensate consists of
a large number of atoms, we introduce the condensate order parameter
or wave function $\Phi_i=\langle\Psi_i\rangle$ for the $i$th component. Neglecting
quantum fluctuations we arrive at the mean field energy functional,
\begin{eqnarray}
{\cal E} &=&\int d\vec r\, \left[ \Phi_i^*\left(-{\hbar^2 \over
2M}\nabla^2 + V_{\rm ext}+E_i\right)
\Phi_i \right.\\
&+&\left. {c_0\over 2}\Phi_i^*\Phi_j^*\Phi_j\Phi_i + {c_2\over 2}
\Phi_k^*\Phi_i^*\left(F_\gamma \right)_{ij}\left(F_\gamma\right)_{kl} \Phi_j
\Phi_l\right], \nonumber \label{e}
\end{eqnarray}
from which the coupled GP equations can be derived according to
$i\hbar\,\partial \Phi_i/\partial t = \delta {\cal E}/\delta
\Phi_i^*$. They are given below in explicit form as
\begin{widetext}
\begin{eqnarray}
i\hbar {\partial \over \partial t}\Phi_+ &=&
    \left[-{\hbar^2\over 2M}\nabla^2+V_{\rm ext} + E_+ + c_0n
    + c_2(n_++n_0-n_-)\right]\Phi_++c_2\Phi_0^2\Phi_-^*,\nonumber\\
i\hbar {\partial \over \partial t}\Phi_0 &=&
    \left[-{\hbar^2\over 2M}\nabla^2+V_{\rm ext} + E_0 + c_0n
    + c_2(n_++n_-)\right]\Phi_0+2c_2\Phi_+\Phi_-\Phi_0^*,\label{3de}\\
i\hbar {\partial \over \partial t}\Phi_- &=&
    \left[-{\hbar^2\over 2M}\nabla^2+V_{\rm ext} + E_- + c_0n
    + c_2(n_-+n_0-n_+)\right]\Phi_-+c_2\Phi_0^2\Phi_+^*,\nonumber
\end{eqnarray}
where $n = \sum_i n_i$ is the total condensate density and $n_i = |\Phi_i|^2$.

The external trap is assumed harmonic $V_{\rm ext} = M(\omega_\perp^2 r_\perp^2
+ \omega_z^2 z^2)/2$ with cylindrical symmetry $\omega_x=\omega_y
=\omega_\perp$, and $\omega_z \ll \omega_\perp$, i.e., cigar shaped.  Following
the successful approach of the NPSE description as for a single component
scalar condensate \cite{Salasnich04}, we factor the wave function into
transversal and longitudinal functions as
\begin{eqnarray}
\Phi_i(\vec r_\perp,z;t) &=& \sqrt N \,\Phi_\perp(\vec r_\perp; \chi(z,t)) f_i(z,t),
\label{az}
\end{eqnarray}
where $\chi$ and $f_i$ are variational functions which depend on $z$ and $t$.
$\Phi_\perp$ is the transversal wave function, satisfying $\int d\vec r_\perp
|\Phi_\perp|^2 = 1$, and is assumed identical for all components.
Substituting Eq. (\ref{az}) into Eq. (\ref{e}), we obtain the
Lagrangian of our system as
\begin{eqnarray}
{\cal L} &=& \int d\vec r \sum_i \Phi_i^*(\vec r,t)\left[i\hbar {\partial \over
\partial t}+{\hbar^2\over 2M}\nabla^2
- V_{\rm ext}-E_i-{c_0N\over 2}\sum_j |\Phi_j|^2\right]\Phi_i(\vec r,t) \nonumber \\
&&- {c_2N\over 2}\int d\vec r
\left[|\Phi_+|^4+|\Phi_-|^4+2|\Phi_+|^2|\Phi_0|^2 +2|\Phi_-|^2|\Phi_0|^2 -
2|\Phi_+|^2|\Phi_-|^2 +
2\Phi_0^{*2}\Phi_+\Phi_-+2\Phi_+^*\Phi_-^*\Phi_0^2\right] \nonumber \\
&=& \int dz \left\{\sum_i f_i^*(z,t)\left[i\hbar {\partial \over
\partial t}+{\hbar^2\over 2M}{\partial^2\over \partial z^2}
- V(z)-E_i-E_\perp(\chi)- \eta(\chi){c_0N\over 2}\rho(z)\right]f_i(z,t) -
\eta(\chi){c_2N\over 2}S_2\right\},
\end{eqnarray}
where $V(z) = M\omega_z^2 z^2/2$. $E_\perp$ is the transverse mode energy, and
$E_\perp(\chi) = \int d\vec r_\perp \Phi_\perp^*[-(\hbar^2\nabla_\perp^2/2M) +
(M\omega_\perp^2 r_\perp^2/2)] \Phi_\perp $. $\eta$ is the scaling factor of
the nonlinear interaction strength, $\eta(\chi) = \int d\vec r_\perp
|\Phi_\perp|^4$, $\rho(z) = \sum_i|f_i|^2$, and $S_2$ is independent of $\chi$
and given by
\begin{eqnarray}
S_2 &=& \left(|f_+|^4+|f_-|^4+2|f_+|^2|f_0|^2 +2|f_-|^2|f_0|^2 -
2|f_+|^2|f_-|^2 + 2f_0^{*2}f_+f_-+2f_+^*f_-^*f_0^2\right). \nonumber
\end{eqnarray}
To obtain the above result, we have also assumed a weak time and $z$ dependence
of the transverse wave function, i.e., $\partial \Phi_\perp /\partial t \simeq
0$ and $\nabla^2 \Phi_\perp \simeq \nabla_\perp^2\Phi_\perp$. The effective
quasi-1D NPSE for a spin-1 condensate can now be derived from the least action
principle of the above Lagrangian,
\begin{eqnarray}
i\hbar {\partial \over \partial t}f_+ &=&
    \left[-{\hbar^2\over 2M}{\partial^2 \over \partial z^2}+ V(z) +E_+ + E_\perp + c_0N \eta
    \rho + c_2N\eta (\rho_++\rho_0-\rho_-)\right]f_++c_2N\eta f_0^2f_-^*,\nonumber\\
i\hbar {\partial \over \partial t}f_0 &=&
    \left[-{\hbar^2\over 2M}{\partial^2 \over \partial z^2}+V(z) +E_0 + E_\perp + c_0N\eta
    \rho + c_2N\eta(\rho_++\rho_-)\right]f_0+2c_2N\eta f_+f_-f_0^*,\nonumber\\
i\hbar {\partial \over \partial t}f_- &=&
    \left[-{\hbar^2\over 2M}{\partial^2 \over \partial z^2}+V(z) +E_-+ E_\perp + c_0N\eta
    \rho + c_2N\eta(\rho_-+\rho_0-\rho_+)\right]f_-+c_2N\eta f_0^2f_+^*,\nonumber \\
\rho {\partial E_\perp \over \partial \chi} &+& \left({c_0N\over 2}\rho^2 +
{c_2N\over 2}S_2 \right){\partial \eta \over \partial \chi} = 0, \label{npse}
\end{eqnarray}
where $\rho=\sum_i\rho_i$ is the total density and $\rho_i=|f_i|^2$
is the density of the $i$th component.
\end{widetext}

We discuss two separate ansatzes for the transverse function applicable
respectively for the cases of weak and strong atomic interactions.
\begin{itemize}
\item \textbf{A Gaussian ansatz}\\
For weak atomic interaction when $\mu - E_\perp \ll \hbar \omega_\perp$
is satisfied, the transverse wave function can be taken as
a Gaussian function of a variable width,
\begin{eqnarray}
\Phi_\perp(\vec r_\perp; \chi(z,t)) &=& {1\over
\pi^{1/2}\chi}\exp[-r_\perp^2/2\chi^2].
\end{eqnarray}
The transverse mode energy and scaling factor are then given by
\begin{eqnarray}
E_\perp &=& {\hbar\omega_\perp\over 2} \left({a_\perp^2\over \chi^2} +
{\chi^2\over a_\perp^2}\right), \\
\eta &=& {1\over 2\pi \chi^2},
\end{eqnarray}
where $a_\perp = \sqrt{\hbar/M\omega_\perp}$.
\item \textbf{A Thomas-Fermi ansatz}\\
For strong atomic interactions when $\mu - E_\perp \gg \hbar \omega_\perp$
holds, the transverse wave function is taken as a Thomas-Fermi (TF) ansatz,
\begin{eqnarray}
&&\Phi_\perp(\vec r_\perp; \chi(z,t)) \nonumber\\
&=& \left\{\begin{array}{cc}\sqrt{2\over
\pi}\; {1\over \chi}\; \sqrt{1-\left({r_\perp \over \chi}\right)^2},&r_\perp\leq \chi;
\\0,&r_\perp > \chi. \end{array}\right.
\end{eqnarray}
The kinetic energy in the transverse direction is neglected, leading to
the transverse mode energy and scaling factor as
\begin{eqnarray}
E_\perp
&=& {\hbar\omega_\perp\over 6}
\left({\chi^2\over a_\perp^2}\right), \\
\eta &=& {4\over 3\pi \chi^2}.
\end{eqnarray}
\end{itemize}

\begin{figure}[h]
\includegraphics[width=3.25in]{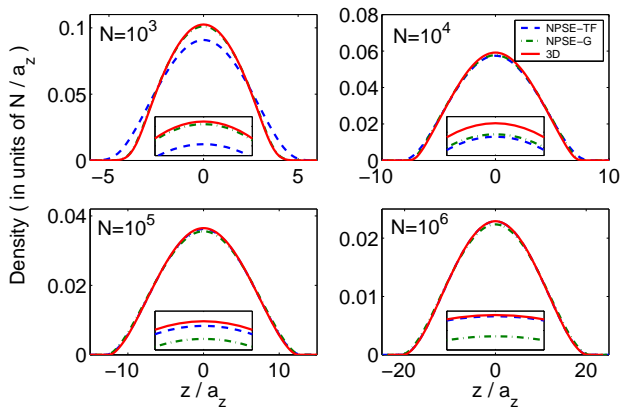}
\caption{The ground state density distribution of the condensate component in
state $|0\rangle$ along the axis of the cigar-shaped trap, for $^{87}$Rb atoms
and without an external magnetic field ($a_z = \sqrt{\hbar/M\omega_z}$). The
inset shows the zoom-in central region. The solid line denotes the ``exact",
while the dashed and dash-dot lines denote respectively the results from our
NPSE with a TF or a Gaussian ansatz for the transverse profile.} \label{s}
\end{figure}

We performed some numerical simulations to illustrate the efficiency and
effectiveness of the NPSE as developed by us for a spin-1 condensate in a
cigar-shaped trap. For the first example, we computed the ground state of a
$^{87}$Rb spin-1 condensate by propagating the GP equations and the effective
1D NPSE with an imaginary time. The atomic parameters of $^{87}$Rb are $a_0 =
101.8\, a_{\rm B}$ and $a_2 = 100.4 \, a_{\rm B}$ \cite{para2}. The trap
frequencies are $\omega_\perp = (2\pi) 240$ Hz and $\omega_z = (2\pi) 24$ Hz.
The ``exact" solution as given by the ground state of the full 3D coupled GP
equations (\ref{3de}) is calibrated by its effective 1D distribution according
to $|f_i(z)|^2 = \int d\vec r_\perp |\Phi_i|^2$. Figure \ref{s} illustrates the
results for several cases of different total number of atoms, $N$. We note that
with increasing $N$, the mean field interaction becomes stronger. For weak
interactions the quasi-1D NPSE with a Gaussian variational ansatz gives a
better result, while for strong interactions the quasi-1D NPSE with a TF ansatz
is a better choice. Here ``better" means the result obtained from an NPSE is
closer to that of the full 3D solution. We also observe that the quasi-1D NPSE
with a TF ansatz gives a lower central density and overestimates the TF radius
in the weak interaction regime, while the quasi-1D NPSE with a Gaussian ansatz
gives a lower central density and a correspondingly larger width in the strong
interaction regime. Over all, it is interesting to point out that the quasi-1D
NPSE with a Gaussian ansatz is not too bad even in the strong interaction
regime.

\begin{figure}
\includegraphics[width=3.25in]{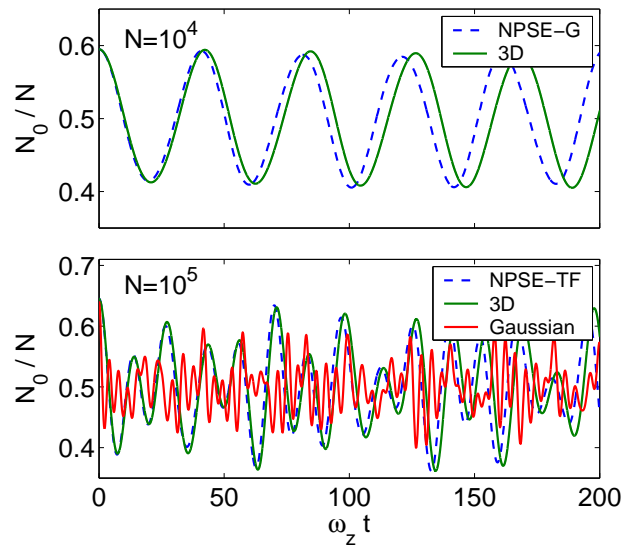}
\caption{The time dependence of the fractional condensate population in the
$|0\rangle$ state $N_0/N$. The thick solid curve denotes the full 3D simulation
while the dashed curve denotes the simulation with our effective quasi-1D NPSE.
We have used $N=10^4$ for the top part and $N=10^5$ for the bottom. As a
comparison we also presented the result obtained from a time-independent
Gaussian ansatz (thin solid curve in the bottom panel), which is shown to give
a poor agreement in the strong interaction regime. } \label{d}
\end{figure}

\begin{figure}
\includegraphics[width=3.25in]{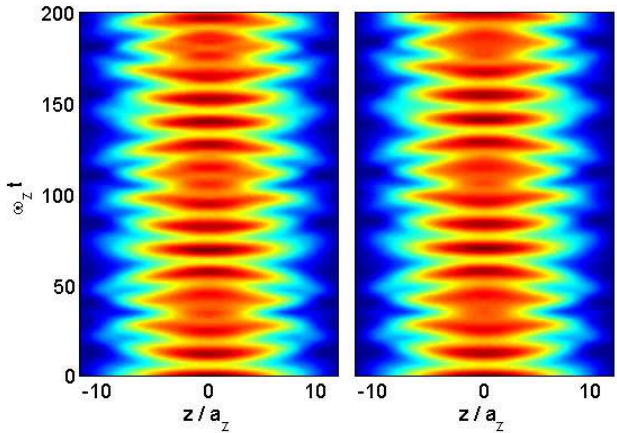}
\caption{The contour plots of the density of the $|0\rangle$ component with
respect to time and z. The left one is from the quasi-1D NPSE while the right
one is the ``exact" result from a full 3D simulation. Parameters are the same
as Fig. \ref{d}.} \label{c}
\end{figure}

To test the quasi-1D NPSE more strictly we study the dynamics of a spin-1
condensate out of equilibrium configuration and compare the results with those
from a full 3D simulation with the coupled GP equations (\ref{3de}). The
initial state is taken as the ground state in a given magnetic field. The
simulation starts after the magnetic field is set zero, and we follow the
spatial-temporal dynamics. With our NPSE, it becomes essentially a trivial
task, and we find that excellent agreements are obtained with a Gaussian
ansatz, e.g., with $N=10^3$ atoms for weak interactions, and a TF ansatz for
strong interactions with $N=10^6$ atoms. In the results to be given below, we
instead use the effective quasi-1D NPSE to simulate the dynamics in the limit
between the strong and weak interactions, i.e., for $\mu - \hbar \omega_\perp
\sim \hbar\omega_\perp$. For the quasi-1D NPSE approach, we use a Gaussian
ansatz with $N=10^4$ atoms (${\cal E} - \hbar \omega_\perp = 1.74
\,\hbar\omega_\perp$)  and a TF ansatz with $N=10^5$ atoms (${\cal E} - \hbar
\omega_\perp = 4.98 \,\hbar \omega_\perp$). Figure \ref{d} displays the time
evolution of the fractional condensate in the $|0\rangle$ state. For weak
interactions, all three components share the space profile along the $z$ axis,
and the out of equilibrium dynamics is periodic \cite{Pu99, Pu00, PRL}. Figure
\ref{d} also clearly shows the periodic motion for $N=10^4$ atoms, although we
do find that the quasi-1D NPSE gives a slightly shorter period than that of the
full 3D simulation. For strong interactions, the apparent spatial profiles of
the three spin components clearly become different, and the out of equilibrium
dynamics also becomes complicated. Yet still, the quasi-1D NPSE simulations
give results very close to the ``exact'' 3D solution, especially in the short
time range. Figure \ref{c} compares the dependence of the density distribution
of the $|0\rangle$ state component on time and space from quasi-1D NPSE and
full 3D simulation. The excellent agreement clearly demonstrates the efficiency
and effectiveness of the quasi-1D NPSE approach, although we do find that it
always seems to give a slightly shorter oscillation period as compared to the
``exact" result.

Before concluding, we hope to discuss the conditions under which our quasi-1D
NPSE description is applicable. For a true 1D condensate which enters the Tonks
gas regime \cite{1DBEC}, our result is obviously not applicable. In the
derivations of the quasi-1D NPSE, we have assumed a weak time- and z-dependence
of the transverse mode. This assumption is valid only for weak excitations of
the condensate such that the excitation in the transverse direction is
negligible. In other words, the wavelength of the excitation is longer than the
transverse size of the condensate. Under this condition, the transverse mode is
reduced to the ground state which is a Gaussian in the weakly interacting limit
and a TF profile in the strongly interacting limit. For spin-1 condensates
widely discussed now, either of $^{23}$Na or $^{87}$Rb atoms, the
spin-dependent excitation is always weak since $c_2$ is two or three orders of
magnitude smaller than $c_0$. The wavelength of the spin wave is thus larger
than the transversal size of the condensate for a cigar-shaped trap
\cite{Myatt97, Stamper-Kurn98, Stenger98, Chang04a, Schmaljohann04,
Kuwamoto04}. Our resulting NPSE model can thus be directly applied. For a
strongly excited system, one has to include the transverse motion, with a more
general approach as developed by Kamchatnov and Shchesnovich in
\cite{Kamchatnov04} and Salasnich {\it et al.} in \cite{Salasnich04a}.

In summary, we have extended the successful effective quasi-1D nonpolynomial
Schr\"odinger equation (NPSE) for a single-component scalar condensate to a
multi-component spin-1 condensate in a cigar-shaped trap. We have demonstrated
its validity with a Gaussian ansatz for the transverse profile in the weak
interaction regime and with a Thomas-Fermi (TF) ansatz in the strong
interaction regime. We have further demonstrated its effectiveness with studies
on both the static (ground state) and dynamic properties of a spin-1 $^{87}$Rb
condensate in a cigar-shape harmonic trap. With the effective quasi-1D NPSE,
simulations for out of equilibrium condensate dynamics become rather efficient,
thus allowing detailed comparisons with the recently observed spatial temporal
dynamics.\\


This work is supported by the NSF.

\end{document}